\documentclass[prd,twocolumn,showpacs,preprintnumbers,amsmath,amssymb]{revtex4}
\usepackage{graphicx}
\begin{document}

\title{Fisher's zeros of quasi-Gaussian densities of states}
\author{A. Denbleyker}
\author{D. Du}
\author{Y. Meurice}
\affiliation{Department of Physics and Astronomy\\ The University of Iowa\\
Iowa City, Iowa 52242, USA }
\author{A. Velytsky}
\affiliation{
Department of Physics and Astronomy, UCLA, Los Angeles, CA 90095-1547, USA}

\date{\today}
\begin{abstract}
We discuss apparent paradoxes regarding the location of the zeros of the partition function in the complex $\beta$ plane (Fisher's zeros) of a pure $SU(2)$ lattice gauge theory in 4 dimensions. 
We propose a new criterion to draw the region of the complex $\beta$ plane where 
reweighting methods can be trusted when the density of states is almost but not exactly Gaussian. We propose new methods to infer the existence of 
zeros outside of this region. We demonstrate the reliability of these proposals with  quasi Gaussian 
Monte Carlo distributions where the locations of the zeros can be calculated by independent numerical methods. The results are presented in such way that the methods 
can be applied for general lattice models. Applications to specific lattice models 
will be discussed in a separate publication. 
\end{abstract}
\maketitle

\section{Introduction}

The locations of the zeros of the partition function in the complex $\beta$ plane contain  
important information regarding the behavior of lattice models and statistical systems at small and large 
$\beta$ but also about possible phase transition or crossover behavior at intermediate values of $\beta$. In the following, $\beta$ can either be $2N/g^2$ for a lattice gauge 
theory or $-J/kT$ for an Ising model or its sigma model generalizations. 
In the following, we will use the notation $S$ 
for the sum over all the plaquettes. For the Ising case, 
this quantity should be replaced by the sum of the product of neighbor spins over all the links.  

The importance of the zeros of the partition in the $\beta$ plane, not to be confused with the 
Lee-Yang zeros in the complex magnetic field plane, was first pointed out by M. Fisher for the two-dimensional Ising model and are called ``Fisher's zeros'' \cite{fisher65}. 
The relationship between the location of the zeros and the order of the 
phase transition has been discussed in Refs. \cite{janke00,janke04}.

In practice, one can calculate the partition function in the complex $\beta$ plane 
up to an overall constant by 
using the reweighting method with respect to a real value $\beta_0$ that is not a zero of the partition function. The partition function at $\beta=\beta _0 
+\Delta \beta$ is proportional to the average of $\exp (-\Delta \beta S)$ 
calculated at $\beta_0$:
\begin{equation}
Z(\beta_0+\Delta \beta)=Z(\beta_0)<\exp (-\Delta \beta S)>_{\beta_0}\ .
	\label{eq:pf}
\end{equation}
As $S$ has a distribution peaked near a value of the order of the number of sites, 
$Z(\beta_0+\Delta \beta)$ has rapid oscillations when the imaginary part of $\Delta \beta$ is increased and it is convenient to subtract $<S>$ from $S$. 
The main quantity that we will calculate is  
\begin{eqnarray}
\label{eq:pf2}
<&\exp& [-\Delta \beta (S-<S>_{\beta_0})]>_{\beta_0}\\ \nonumber
= &\exp& [\Delta \beta <S>_{\beta_0}]Z(\beta_0+\Delta \beta)/Z(\beta_0)\ .
\end{eqnarray}
It has the same complex zeros as $Z(\beta_0+\Delta \beta)$.

In lattice gauge theory (LGT), the Fisher's zeros of pure (no quarks) $SU(3)$ gauge theories 
with a Wilson action on $2\times L^3$ and 
$4\times L^3$ lattices were studied using Monte Carlo (MC) configurations to 
perform the average, in order to locate the finite temperature phase transition 
\cite{alves90,alves90b,alves91}. 
At zero temperature, no phase transition is expected between small and large $\beta$ and consequently, the complex zeros should stay away from the real axis in the infinite 
volume limit. 

Recently, it has been suggested \cite{third} that a pair of complex conjugated zeros very close to the real axis 
could be responsible for the temporary behavior of the perturbative series 
for the average plaquette indicating 
a finite radius of convergence (power growth) rather than the expected factorial growth.

The existence of a zero near $\beta =5.55+i0.12$ on a $4^4$ lattice \cite{alves90b,alves91} was confirmed in Ref. \cite{lattice06}. 
However, it seemed that for a $8^4$ lattice, the zeros would be far away from 
the region of confidence defined in Ref. \cite{alves91}. 
Using the same criterion, it is also not clear that the estimation \cite{falcioni82} of a $SU(2)$ zero 
near $\beta=2.23+i0.155$ on a $4^4$ lattice is reliable. These issues are reviewed in Sec. \ref{sec:su2}.

An important feature of the $SU(2)$ distributions of actions is that they are almost Gaussian. The distribution can be expressed as the product of a density of states $n(S)$ 
(also called spectral density) 
and the Boltzmann weight $\exp(-\beta S)$. Consequently, a quasi-Gaussian distribution 
corresponds to a quasi-Gaussian density of states. 
In the Gaussian limit, the partition function has no zeros \cite{alves91}.

The main goal of the article is 
to provide new methods to find the zeros of the partition function in terms 
of the small deviations from the Gaussian behavior. In order to be able to apply 
the methods to generic lattice models, we first introduce natural units (Sec. \ref{sec:natural}) where the mean and the variance are reduced to 0 and 1. 
It is important to understand that for lattice models, the variance of $S$ increases with the volume like $V^{1/2}$ and 
that when we work in natural units, we need to go to larger imaginary part when the volume increases in order to study the behavior near a fixed value of $\beta$. 
To make this concrete, in 4 dimensions, if we double the linear size of the lattice, 
the radius of confidence shrinks by a factor 4. 

The rest of the paper is organized as follows. The Gaussian case and the notion of circle of confidence are reviewed in Sec. \ref{sec:gaussian}. Possible causes for 
underestimation of the fluctuations are discussed in Sec. \ref{sec:error}. 
A new criterion designed to complement the notion of circle of confidence for 
non-Gaussian distributions is proposed in Sec. \ref{sec:new} and applied to two examples 
in Sec. \ref{sec:new}. The conjecture that the real part of the zero is located 
at a value of $\beta$ where the variance of $S$ is maximal is verified for these examples in Sec. \ref{sec:conjecture}. Fitting methods to locate the zeros in region 
not accessible by MC methods are proposed in Sec. \ref{sec:indirect} and 
proved successful and consistent for the two examples considered. 
\section{The case of $SU(2)$}
\label{sec:su2}
\begin{figure}
\includegraphics[width=2.3in,angle=270]{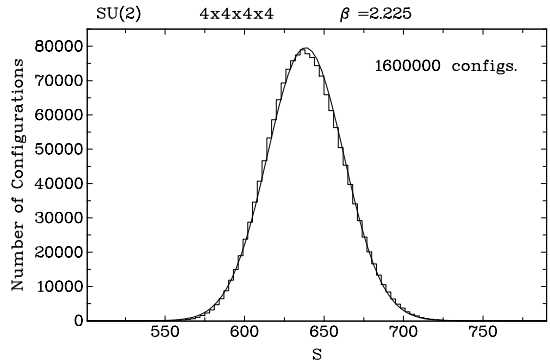}
\caption{Distribution of the 1,600,000 values of $S$ in an histogram with 100 bins 
for an $SU(2)$ pure gauge theory on a $4^4$ lattice at $\beta =2.225$  .}
\label{fig:su2hist}
\end{figure}

For $SU(2)$ LGT in 4 dimensions, action distributions are nearly Gaussian. This is illustrated in Fig. \ref{fig:su2hist}. However small departures from the Gaussian behavior can be observed. 
Groups of neighbor bins move above and below the Gaussian curve at a level that 
cannot be explained by statistical fluctuations.  
It is well-known \cite{alves91} that Gaussian distributions have no complex zeros. 
Consequently, the small departures should be the cause of possible zeros. 
In Ref. \cite{falcioni82}, a zero was reported near $\beta = 2.23+i0.155$ using a 
reweighting of a MC distribution obtained for $\beta_0=2.225$ and 39,000 configurations.
This result was obtained by locating the intersection of the zeros of the real 
and imaginary part. An error of 0.01 on the real and imaginary part was quoted.

Later, a criterion was proposed \cite{alves91} to determine a region of confidence for the 
zeros. The criterion is based on the Gaussian approximation and can be summarized as follows. We first define 
\begin{equation}
	\sigma_S^2=<S^2>-<S>^2\ ,
\end{equation}
Using Gaussian integration it can be shown that for $N_{conf.}$ independent configurations, 
the fluctuation in $\exp (-\Delta\beta (S-<S>))$ become of the same size as the average for 
\begin{equation}
	|\Delta\beta|^2< \ln (N_{conf.})/\sigma_S^2 \ .
\end{equation}
This defines a circle of confidence with radius $\sqrt{\ln (N_{conf.})/\sigma_S^2}$.
For the calculation of Ref. \cite{falcioni82}, 
$N_{conf}=39,000$ and $ \sigma_S^2=580$, we obtain a radius of confidence of 0.135.
Consequently an imaginary part of 0.155 is definitely outside the circle of 
confidence. This could either mean that there are larger errors than the ones quoted or that the Gaussian criterion does not apply precisely.

We have attempted to reproduce the result of Ref. \cite{falcioni82}
using the heatbath method of Kennedy and Pendleton \cite{kennedy85} to generate 1,600,000 configurations. We have then used various methods to produce 39,000 decorrelated 
configurations. As explained in the Introduction, in order to eliminate rapid oscillations we subtracted $<S>$ in the 
exponent. This does not affect the zeros of the partition function. In Ref. \cite{falcioni82}, this subtraction was not performed 
and the zero level curves of the real part or the imaginary part appear with an approximate spacing of $\pi/<S>\simeq 0.005$ in the imaginary direction. 

\begin{figure}
\includegraphics[width=0.8\columnwidth]{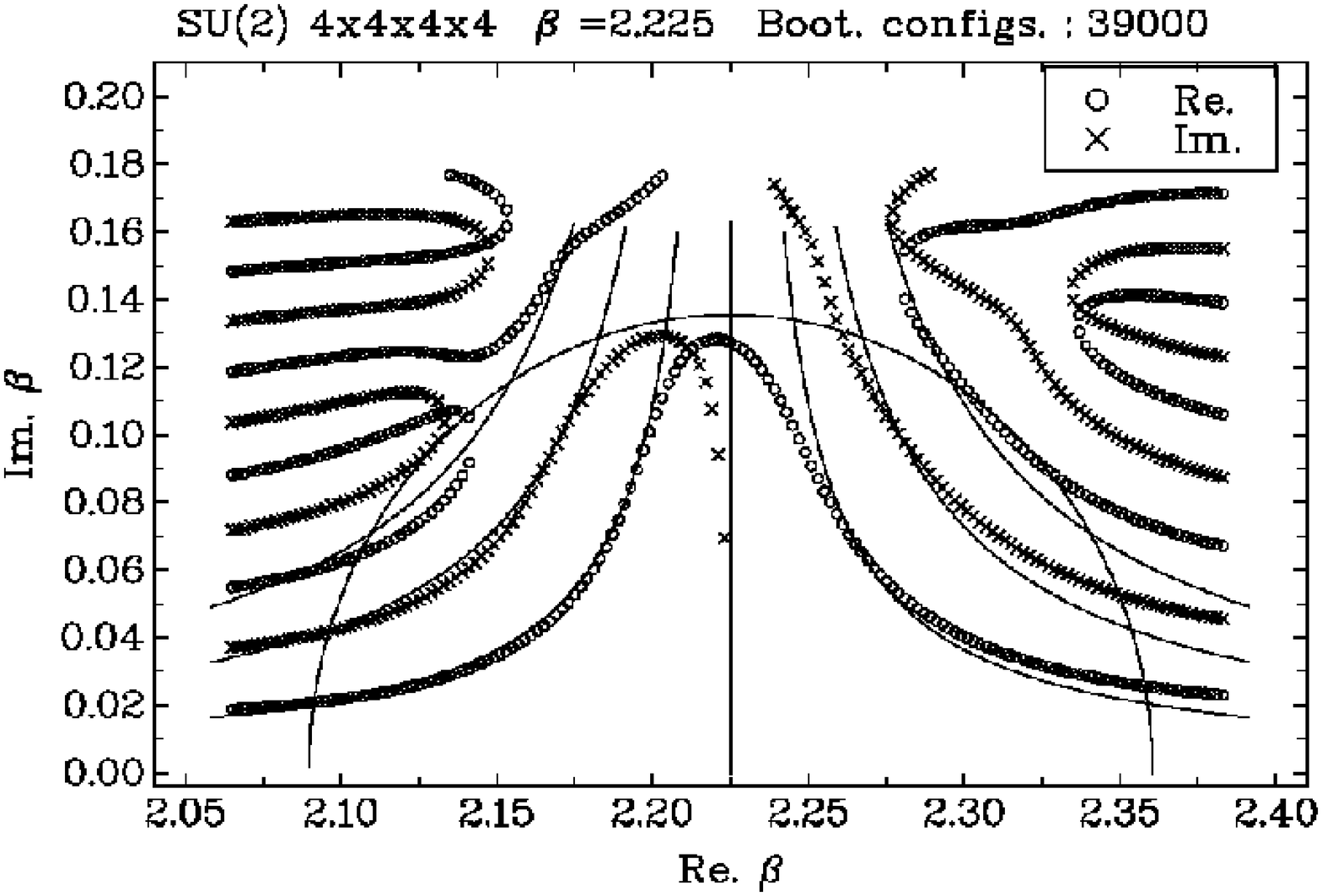}
\vskip-20pt
\includegraphics[width=0.8\columnwidth]{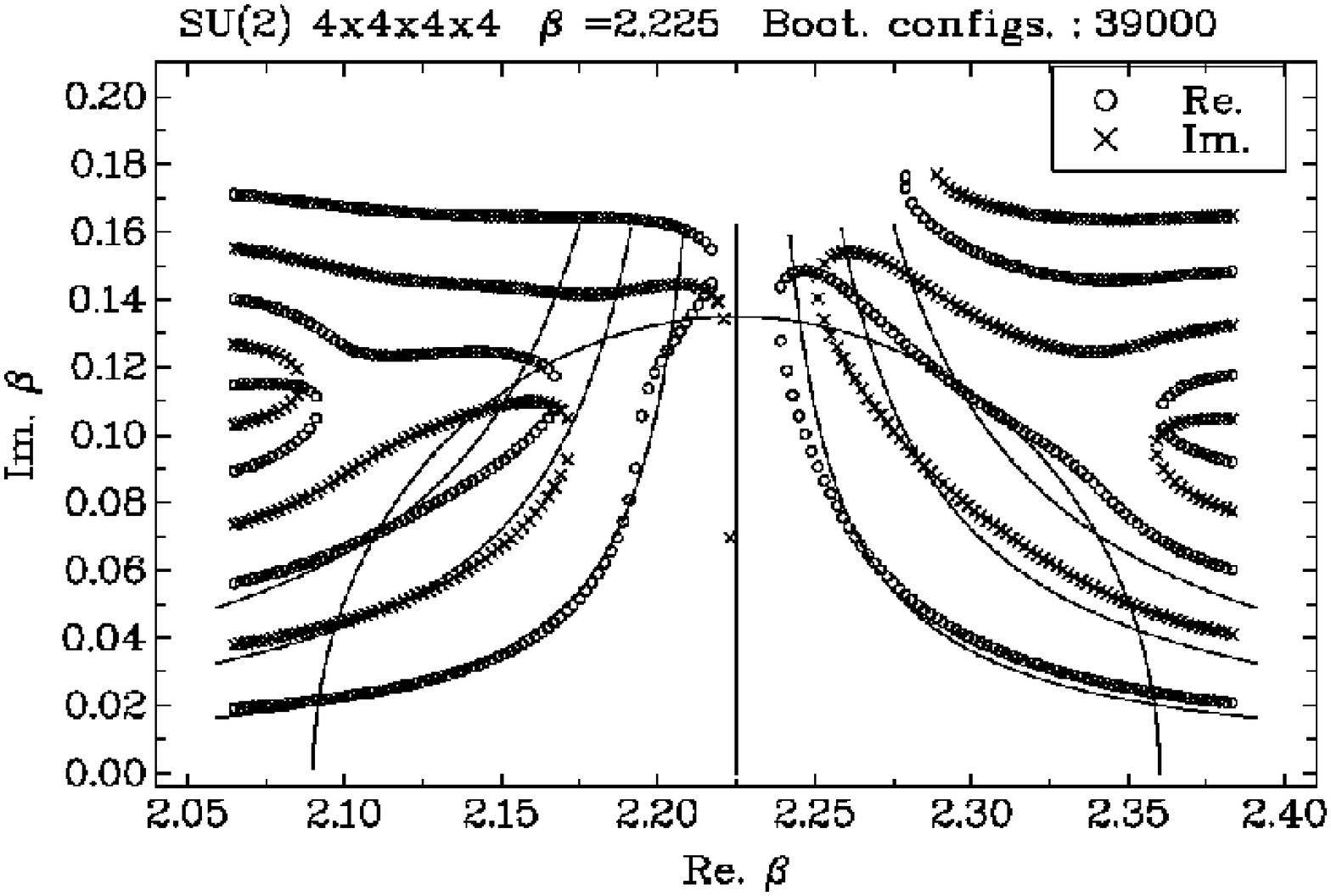}
\caption{Zeros of the real (circles) and imaginary (crosses) part for 
39,000 configurations of a $SU(2)$ gauge theory on a $4^4$ lattice with
 $\beta_0=2.225$.
The solid lines are the circle of confidence and the hyperbolas of the Gaussian approximation described in the text.}
\label{fig:su2zeros}
\end{figure}

We have not found clear evidence for the result of Ref. \cite{falcioni82}. 
In Fig. \ref{fig:su2zeros} we show the zeros of the real and imaginary part obtained from two independent 
bootstrap procedures where the configurations are picked at random with possible repetitions from the large original set of 1,600,000. 

In both graphs, no zeros are found within the error bars of the Ref.  \cite{falcioni82}. 
In addition, zeros are found slightly within the circle of confidence and at different 
locations. 
For comparison, the figures show the location of the zeros of the real and the 
imaginary part that would be seen if the distribution was a Gaussian with the 
same mean and variance than the sample. 
These zeros appear on non-intersecting hyperbolas \cite{alves91} given by the 
formula
\begin{equation}
Im \beta \Delta Re \beta= \pi n/\sigma_S^2\ ,
\end{equation}
with odd (even) $n$ standing for the zero of the real (imaginary) part. 
One robust feature found in the collection of figures produced with 
independent bootstraps is the way the MC data departs from these hyperbolas. 
On the far right the MC data is above the hyperbolas and then goes below as we move toward the center. On the left, the MC data follows more closely the  Gaussian 
hyperbolas. In summary, the MC data seems slightly tilted to the left. 
Another robust feature which seems correlated with this left tilt is that the complex 
zeros within the Gaussian circle of confidence appear on the left side of the 
figure.

At this point, it is not clear to us that with the existing information it is possible to 
draw solid conclusions regarding the existence, the location or the absence of 
complex zeros in the portion of the complex $\beta$ plane considered.
Consequently, we decided to study examples of quasi Gaussian distributions 
where the location of the zeros can be determined by independent numerical methods.
Another remark, understanding the reweighting in the complex plane may also be important for problems with a chemical potential.

\section{Natural units and Notations}
\label{sec:natural}
In the following we will consider MC distributions generated for the purpose of testing 
general methods that can be applied for a broad range of models. In order to translate 
the results presented hereafter in a form that is useful for specific models, we 
will redefine the origin and the scale of $S$.  

We have seen in the previous section, that the distribution of Wilson action $S$
is approximately Gaussian. Given this distribution, it is always possible to take 
every value of $S$, subtract 
the mean and then divide by the variance. We then obtain a distribution which 
is approximately a normal Gaussian. The advantage of using this normal form 
is that the results that follow can be used for other models, for instance scalar 
models or spin models where the typical scales are completely different. 
We define a reduced form:  
\begin{equation}
S_{red.}=(S-<S>)/\sigma_S\ .
\label{eq:can}
\end{equation}
As $<S>$ is already subtracted in the average considered in Eq. (\ref{eq:pf2}), we only
need to rescale $\Delta\beta$ in order to eliminate the variance:
\begin{equation}
\beta_{red.}=\Delta \beta \sigma _S\ .
\end{equation}
For instance, in Fig. \ref{fig:su2zeros}, the scale of the figure is about 
0.16. In the reduced coordinates, the size of the figure would be about 3.85 and could be compared with corresponding graphs for other models.

In order to make some equations more readable we introduce the following notations:
\begin{eqnarray}
\label{eq:notations}
\beta_{red.} &\equiv& x+i y  \nonumber \\
f&\equiv& <\exp (-\beta_{red.} S_{red.})> \nonumber \\
R&\equiv& Re f\nonumber \\
I&\equiv& Im f \nonumber \\
\sigma^2_{Re}&\equiv& < (Re \exp (-\beta_{red.} S_{red.})-R)^2>\nonumber \\ \nonumber
\sigma^2_{Im}&\equiv&< (Im\exp (-\beta_{red.} S_{red.})-I)^2>\nonumber \\ \nonumber
\sigma^2_f&\equiv&\sigma^2_{Re}+\sigma^2_{Im}\\ \nonumber
\end{eqnarray}
It is understood that $f$, $R \dots$ etc are all functions of $\beta_{red.}$. 
In the following, unless specified, we will work mostly with $S_{red.}$ and $\beta_{red.}$ and the 
subscripts ``red.'' will be dropped.

\section{The gaussian case revisited}
\label{sec:gaussian}

\subsection{Data generation}

\begin{figure}
\includegraphics[width=2.6in,angle=270]{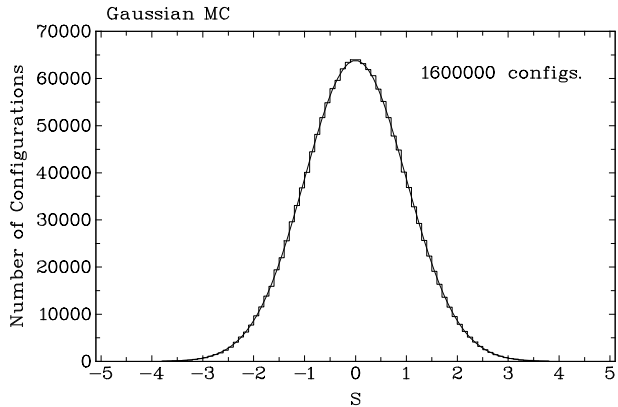}
\caption{Distribution of the 1,600,000 values of $S$ in 100 bins (above) and the 
corresponding $r_i$ (below).}
\label{fig:full}
\end{figure}

We used the Metropolis algorithm in order to generate 
values of $S$ with a normal distribution:
\begin{equation}
\label{eq:normal}
	P(S)=(2\pi)^{-1/2} \exp (-S^2/2)\ .
\end{equation}
The upgrade $S\rightarrow S+ \delta S$ were performed with $\delta S$ uniformly distributed in an interval $[-R,R]$. 
The probability of accepting the change (acceptance) is 
\begin{eqnarray}
\label{eq:acc}
	&1&-(1/2R)(2\pi)^{-1/2}\int_{-\infty}^{+\infty}dS\int_{-R}^R d\delta S 
	\exp (-S^2/2) \nonumber \\
&\ &	[\theta(\delta S)\theta (2S+\delta S)+\theta(-\delta S)\theta 
	(-2S-\delta S)] \ .
\end{eqnarray}
We have generated 1,600,000 values of $S$ for $R=1$. The average of this set of 
values was 0.00087 and the acceptance was 0.8047 in good agreement with Eq. (\ref{eq:acc}) which predicts 0.804583. The values were sorted in 100 bins of equal 
width as shown in Fig. \ref{fig:full}. The departures from the Gaussian behavior 
are difficult to see.
Before analyzing these departures, we will first address the question of autocorrelations 
of successive configurations.
\subsection{Autocorrelations}
\begin{figure}
\includegraphics[width=2.6in,angle=270]{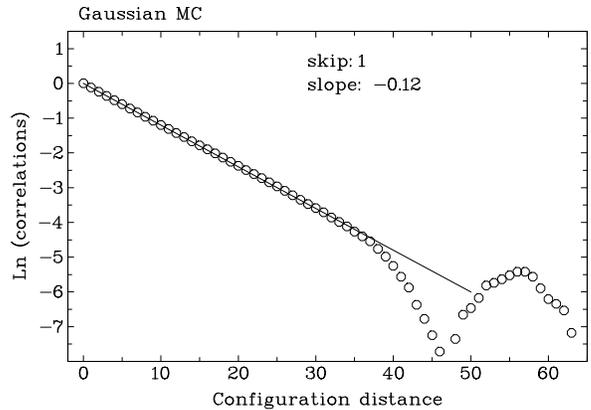}
\caption{ln($|C_l|$) versus $l$ for the original set of 1,600,000 values.} 
\label{fig:fullcorr}
\end{figure}

\begin{figure}
\includegraphics[width=2.6in,angle=270]{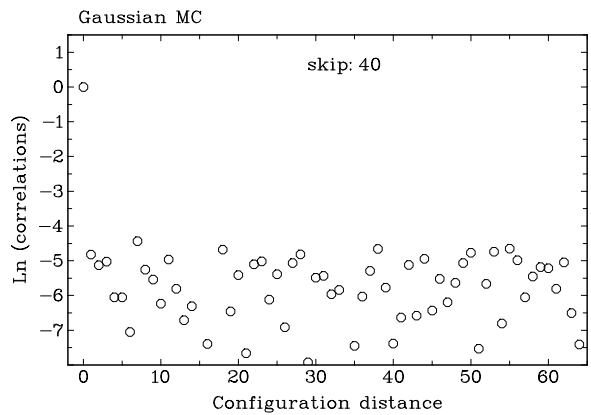}
\caption{ln($|C_l|$) versus $l$ for the subset $S_1, \ S_{41}, \dots$ (above).}
\label{fig:corr40}
\end{figure}

The autocorrelations are defined as usual by the formula
\begin{equation}
	C_l=(1/n)\sum_{m=1}^{n}(S_m-<S>)(S_{m+l}-<S>)\ ,
\end{equation}
where the subscript refers to the configuration number. 
The numerical values are shown on a logarithmic scale in Fig. \ref{fig:fullcorr}. 
The linear part of the graph stabilizes when $n$ is large enough (here about 200,000). 
The slope of the linear part is about -0.12 which means that the correlation time is 
about 8. In order to select reasonably decorrelated values it is necessary to select 
values that are several correlation times apart. In the following, we constructed  subsamples of size 40,000 (about the same size as in \cite{falcioni82}), obtained by selecting one in every 40 values from the original sequence. 
As shown in Fig. \ref{fig:corr40}, the subset show no apparent autocorrelations. 
Note that results similar to the ones obtained in the rest of this section 
can be obtained by using the bootstrap method instead of the skipping method.


\subsection{Histogram analysis for decorrelated data}
\label{subsec:resdec}

\begin{figure}
\includegraphics[width=2.6in,angle=270]{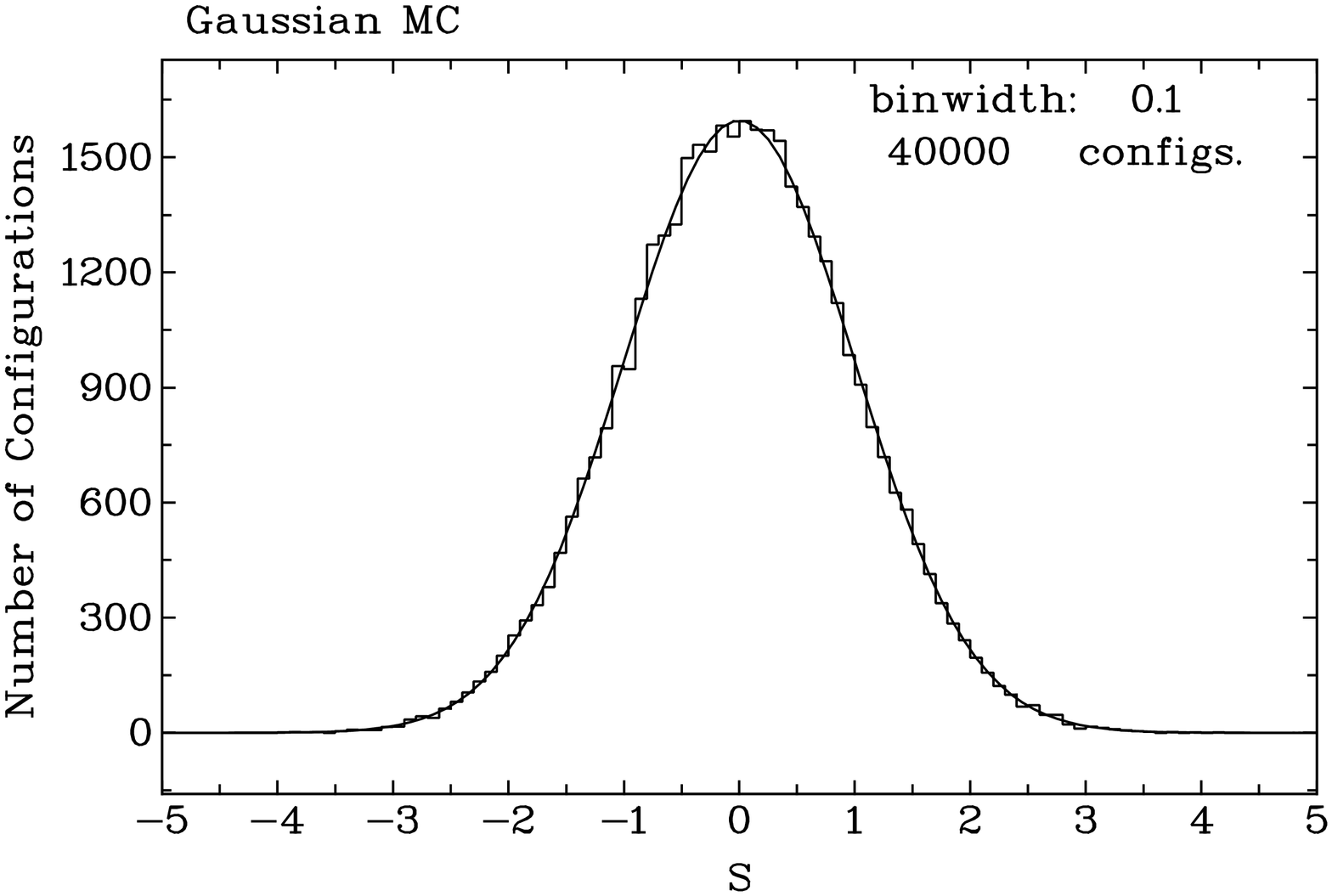}
\vskip-20pt
\includegraphics[width=2.6in,angle=270]{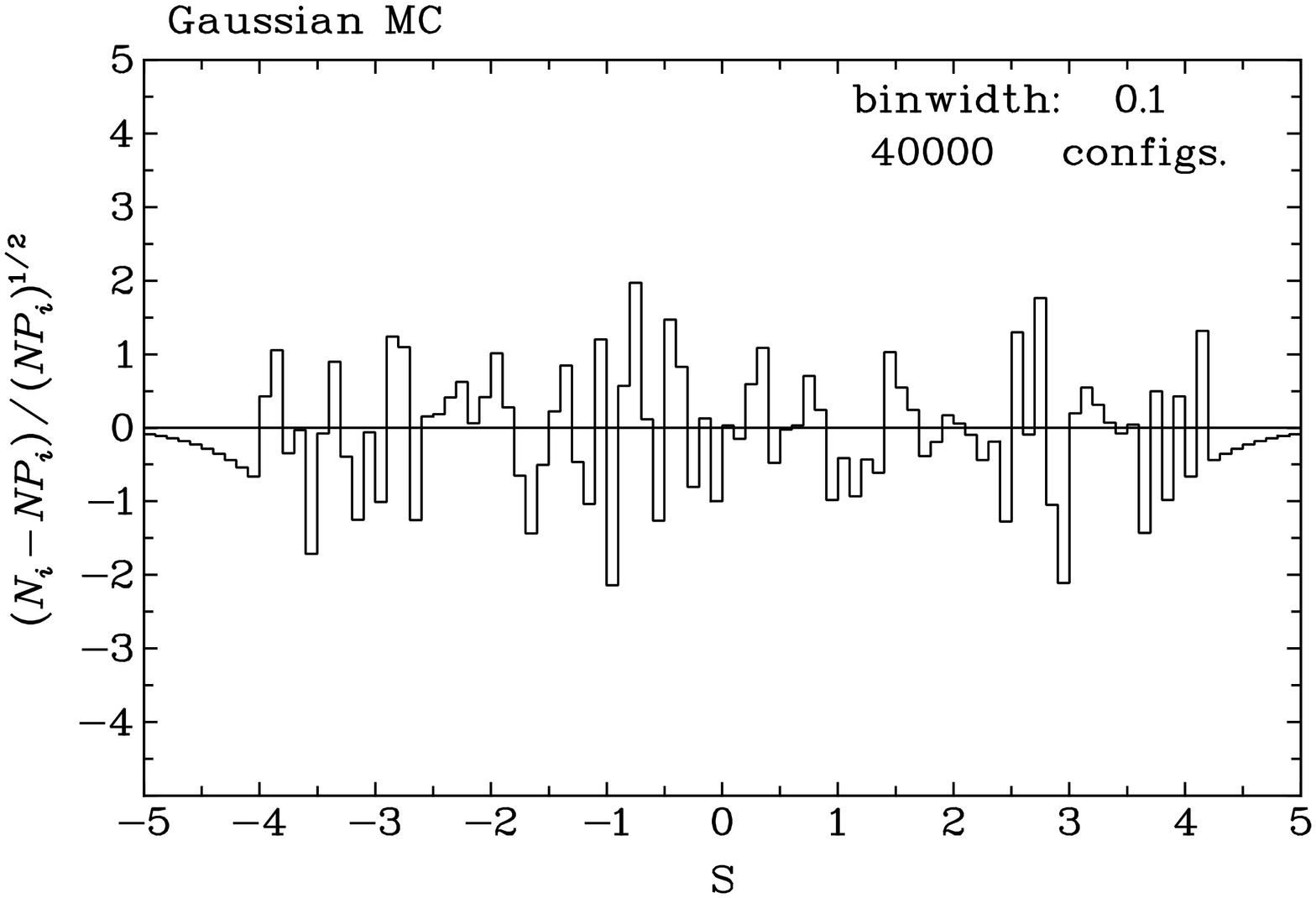}
\caption{Distribution of the 40,000 values of $S_1,\ S_{41},\dots$ in 100 bins (above) and the 
corresponding $r_i$ (below).}
\label{fig:skip40hist}
\end{figure}

The analog of Fig. \ref{fig:full} for the subsample described above is shown in Fig. \ref{fig:skip40hist}.
More information can be gathered by displaying the differences 
in units of the expected fluctuations 
$r_i$ for each bin, namely
\begin{equation}
r_i=(N_i-NP_i)/\sqrt{NP_i} \ ,
\end{equation}
where $N$ is the total number of values (40,000), $P_i$, the expected probability to be in the $i$-th bin, the probability density (for a normal Gaussian) in the middle of the bin multiplied by the bin width, and $N_i$ the number of values in the $i$-th bin. If the $N$ processes were 
independent, we would have a binomial distribution and the average number in the $i$-th bin would be $NP_i$ with a variance 
$\sigma_i^2=NP_i(1-P_i)\simeq NP_i$, since the $P_i$ are small. We expect $|r_i|\sim 1$. 
For the subset obtained with the skipping method  $\sum r_i^2 \simeq 62$ (for 100 bins).

It is important to notice that the $r_i$ show almost no coherent structure. 
The difference tend to have opposite signs and no simple pattern 
seems to be visible except for the tails of the distributions. Coherent negative 
difference are seen for $|S|>4$. This feature has a very simple explanation:  
there are almost no values for $|S|>4$ and we see merely minus the square root of the Gaussian distribution. 
Given that we only only use 40,000 configurations, large values of $|S|$ corresponds to 
small statistics. 
The probability of having $|S|>4$ is about $6\times 10^{-5}$ for a normal distribution.
Consequently, we would expect 2 values in this range in a typical MC run described above. In Fig. \ref{fig:skip40hist}, one value near 4.3 produces a positive spike. 

\subsection{Zeros of $<\exp (-\beta S)>$}

\begin{figure}
\includegraphics[width=0.8\columnwidth]{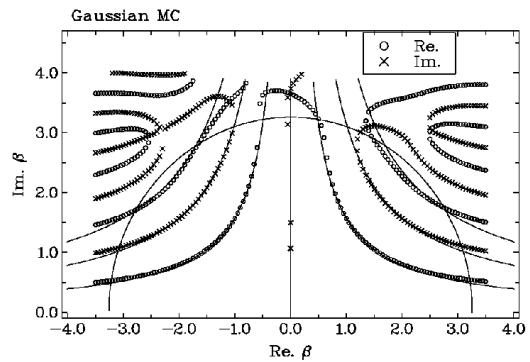}
\caption{Zeros of the real (circles) and imaginary (crosses) part for 
40,000 Gaussian configurations.
The solid lines are the circle of confidence and the hyperbolas of the Gaussian approximation described in the text.}
\label{fig:gausszeros}
\end{figure}

Having made a reasonable effort to produce samples of 40,000 values of $S$ 
that are distributed like a normal Gaussian, we now consider the zeros. 
The zeros of the real and imaginary part of $<\exp (-\beta S)>$ are displayed in 
Fig. \ref{fig:gausszeros} for 40,000 configurations. For comparison, we also show the 
hyperbolas $Im \beta Re \beta = n\pi$ that should be followed by these level curves and 
the circle of confidence $(Re \beta)^2+(Im \beta)^2=\ln(40,000)\simeq (3.26)^2$. 
We see that as we follow the level curves from the inside of the circle toward the boundary, the agreement slowly deteriorates. The apparent complex zero closest to the origin is near $\beta \simeq 1.5+i 3.0$ and is slightly outside the circle of confidence. Fig. \ref{fig:gausszeros} is thus consistent with the known 
absence of zeros in the Gaussian case. Note that it is not impossible to find random samples of 40,000 configurations with complex zeros slightly inside the circle of 
confidence, but this is rather uncommon. By design \cite{alves91},  this should occur in 
16 percent of the cases in average. 

\section{Error analysis for the Gaussian model}
\label{sec:error}

As already shown in Ref. \cite{alves91}, all the quantities of interest can be calculated using standard methods of Gaussian integration.
Remembering the notations introduced in Eq. (\ref{eq:notations}), 
The average and variance of the real and imaginary part can be expressed as:

\begin{eqnarray}
\label{eq:gav}
I&\equiv& \exp((1/2)(x^2-y^2))\cos(xy)\nonumber \\
R&\equiv&\exp((1/2)(x^2-y^2))\sin(xy)\nonumber \\
\sigma^2_{Re}&\equiv& (1/2)\exp(2x^2)(\exp(-2y^2) \cos(4xy)+1)\nonumber \\
&\ &-\exp(x^2-y^2)\cos^2(xy)\\
\sigma^2_{Im}&\equiv& (1/2)\exp(2x^2)(-\exp(-2y^2) \cos(4xy)+1)\nonumber \\
&\ & -\exp(x^2-y^2)\sin^2(xy)\nonumber \\ \nonumber
\end{eqnarray}

We have compared these analytical formulas with MC results for $x=Re\beta$= 1, 2 and 3. 
The errors bars on the MC data point have been estimated by using the MC estimator of the variance divided by $\sqrt{40,000}$.

\begin{figure}
\includegraphics[width=0.8\columnwidth]{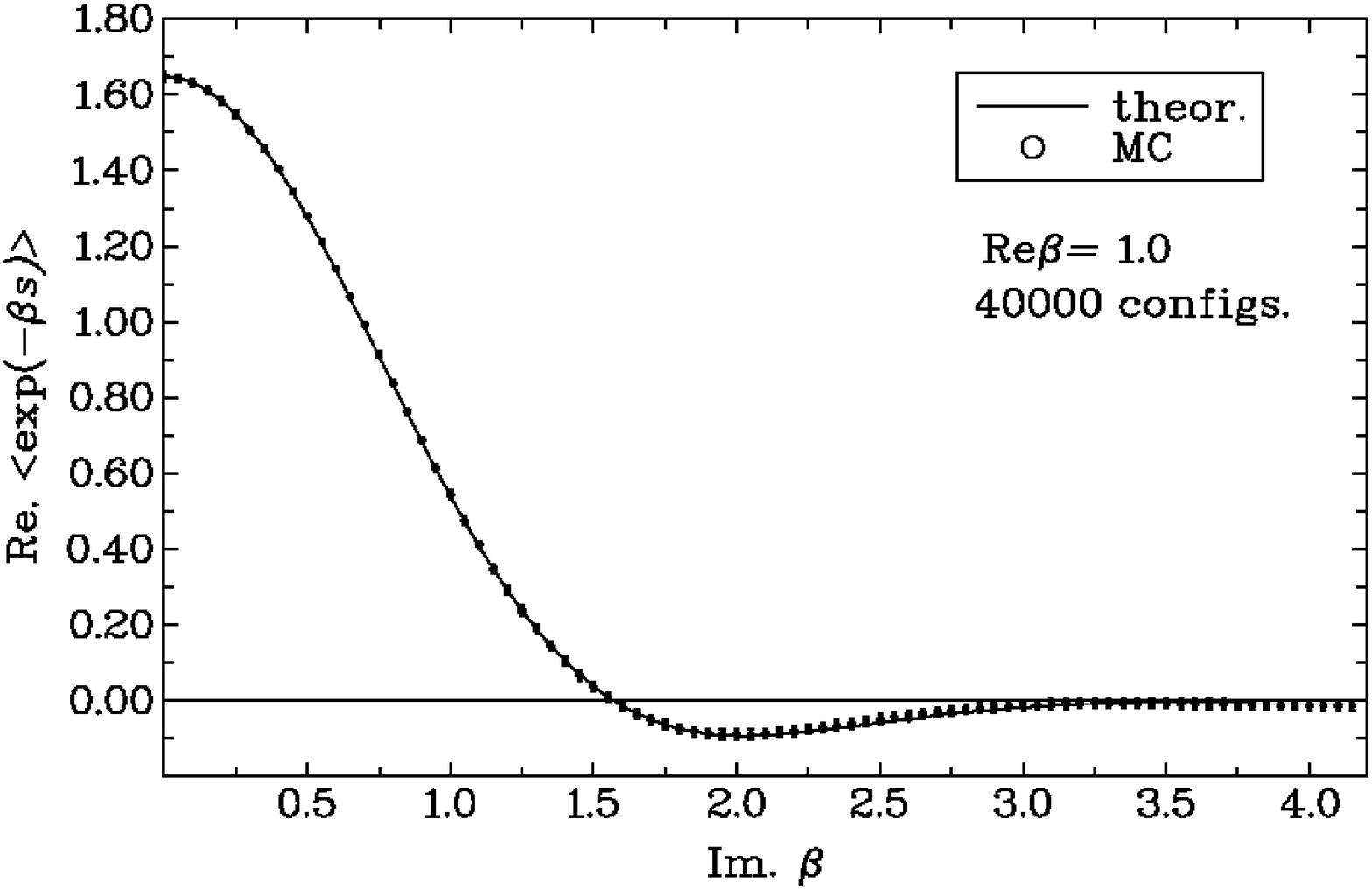}
\vskip-20pt
\includegraphics[width=2.6in,angle=270]{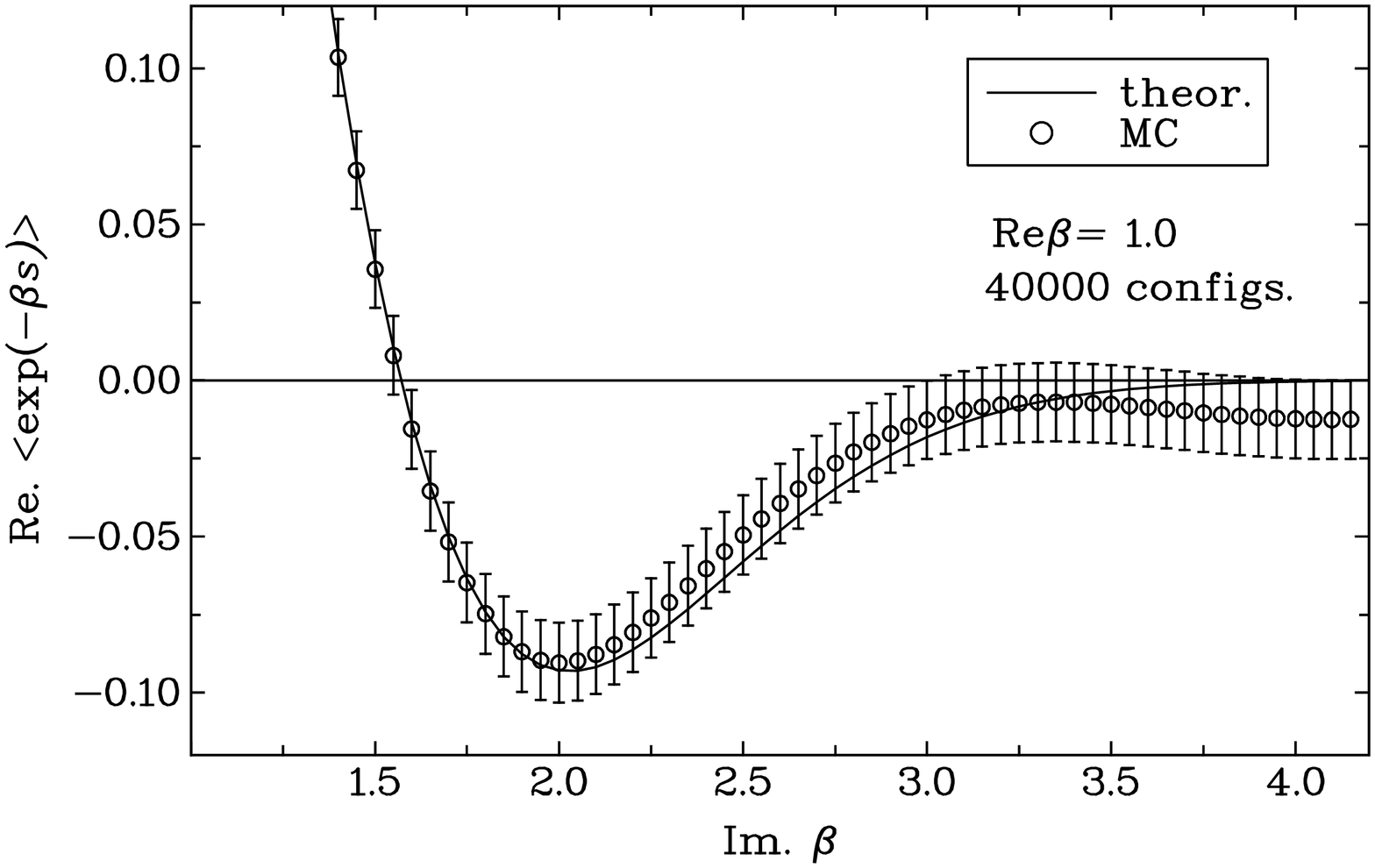}
\caption{$R=Re<\exp(-\beta S>$ versus $y=Im\beta$ at fixed $x=Re\beta$= 1. The continuous line is Eq. (\ref{eq:gav}), the circles are MC results. The lower figure is a 
magnified part of the upper figure.}
\label{fig:re1}
\end{figure}

The results for the real part at fixed $x$ =1 are displayed in Fig. \ref{fig:re1}.  
One sees that the there is a very good agreement until the function becomes smaller than the error bars. This occurs near $y=3$ which is approximately where the crossing of 
the circle of confidence occurs. A more detailed investigation also shows that for this 
value of $x$, the MC estimation of the variance is in good agreement with Eq. (\ref{eq:gav}) within the circle of confidence.

\begin{figure}
\includegraphics[width=0.8\columnwidth]{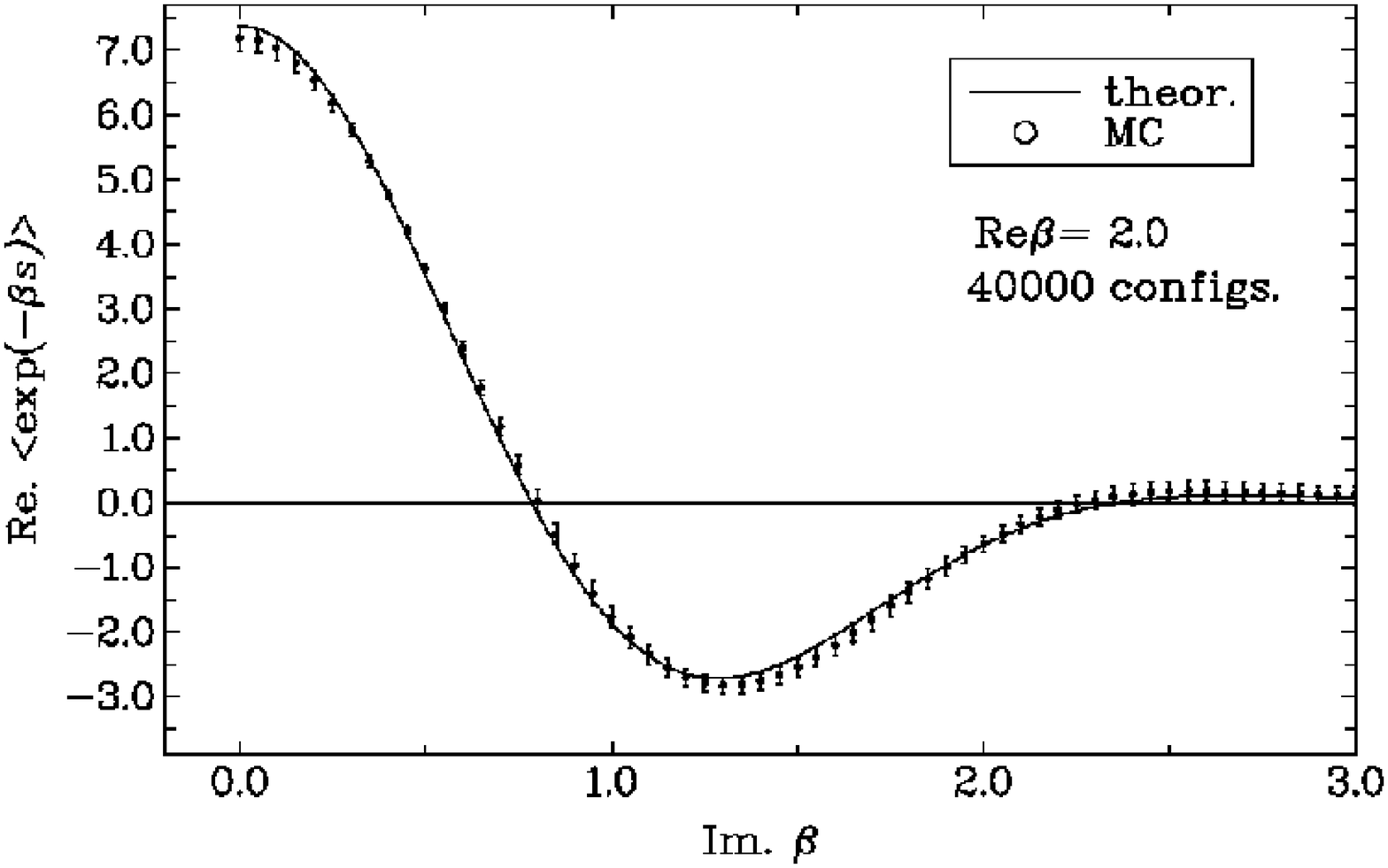}
\caption{$R=Re<\exp(-\beta S>$ versus $y=Im\beta$ at fixed $x=Re\beta$= 2. The continuous line is Eq. (\ref{eq:gav}), the circles are MC results. }
\label{fig:re2}
\end{figure}

\begin{figure}
\includegraphics[width=2.6in,angle=270]{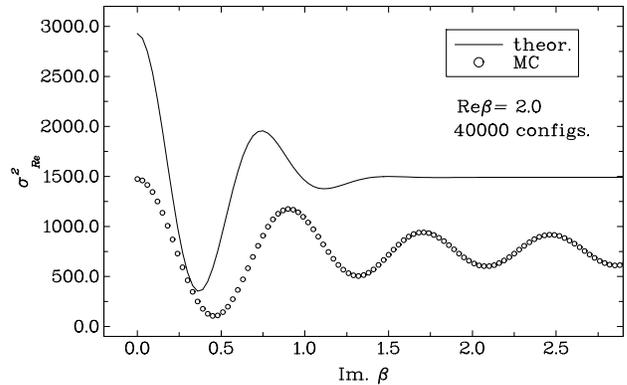}
\caption{ $\sigma^2_{Re}$ versus $y=Im\beta$ at fixed $x=Re\beta$= 2. The continuous line is Eq. (\ref{eq:gav}), the circles are MC results.}
\label{fig:var2}
\end{figure}

\begin{figure}
\includegraphics[width=2.6in,angle=270]{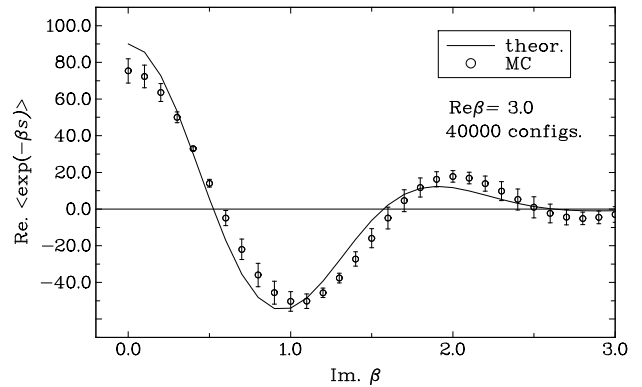}
\caption{$R=Re<\exp(-\beta S>$ versus $y=Im\beta$ at fixed $x=Re\beta$= 3. The continuous line is Eq. (\ref{eq:gav}), the circles are MC results.}
\label{fig:re3}
\end{figure}

\begin{figure}
\includegraphics[width=2.6in,angle=270]{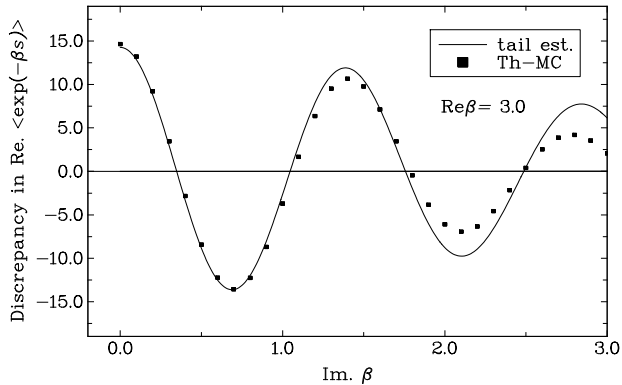}
\caption{Discrepancy between Eq (\ref{eq:gav}) and MC in Fig. \ref{fig:re3} (solid square) and analytical contribution of $S<-4$ to $R=Re<\exp(-\beta S>$. }
\label{fig:distail}
\end{figure}

Larger discrepancies and larger error bars can be seen in Fig. \ref{fig:re2} for $x=2$.
At some places the analytical expression is almost outside the MC error bars. 
Also, the MC estimate of the variance is now significantly lower than the analytical one 
as shown in Fig. \ref{fig:var2}. In other words, the errors bars of Fig. \ref{fig:re2} should be larger.
All the features observed at $x=2$ are dramatically amplified for $x=3$. The analytical expressions is now mostly outside of the MC error bars in Fig. \ref{fig:re3}. A more detailed study shows that 
the error bars are severely underestimated (sometimes by an order of magnitude). 
The discrepancies observed in Fig. \ref{fig:re3} have a simple explanation. 
When $Re\beta$ becomes large, the configurations with large negative values of $S$ 
become very important. We have seen in Sec. \ref{subsec:resdec} that no configurations 
with $S<-4$ were present (while we should typically expect one). 
Fig. \ref{fig:distail} shows that the contribution of the tail $S<-4$ fits very precisely the discrepancy in the average value. The absence of configurations with 
$S<-4$ also explains the underestimation of the error bars.

In summary, when $Re \beta$ is not too large, the MC results are in good agreement 
with Eq. (\ref{eq:gav}) within the circle of confidence. 
When $|Re \beta|$ becomes too large, we probe the tails of the 
$S$ distribution where we have low statistics and obtain unreliable results even 
within the circle of confidence. 

A solution to the problem of low statistics consists in introducing a bias $\exp(-bS)$ in order to probe better the tails of the distribution. In terms of the original 
problems discussed in the introduction this consists in using reweighting for different 
but nearby values of $\beta_0$. This amounts of ``patching" the various data as explained in Refs. \cite{alves89,alves91}. The question will be discussed in Sec. 
\ref{sec:conjecture}.

\section{A new criterion for the region of confidence}
\label{sec:new}

We have seen in the previous sections on the Gaussian distribution, that provided that $Re \beta$ is not too large, 
the MC results are reliable within the circle of confidence and that it is uncommon to 
produce complex zeros in this region. On the other hand for $SU(2)$, complex zeros within 
the circle of confidence are very common. 
Since in this case the distribution is not exactly Gaussian, complex zeros are possible 
but the changes observed in Fig. \ref{fig:su2zeros} suggest that the complex 
zeros seen in the circle of confidence (calculated in the Gaussian approximation) are 
numerical artifacts. 
In general, we may obtain false complex zeros 
if the zero level curves of the real and imaginary part move too far from their exact 
trajectories.  In this section, we propose a criterion to decide when it is the case. 

The Gaussian circle of confidence in the complex $\beta$ plane is defined by the condition 
\begin{equation}
\sigma _f(\beta)/\sqrt{N_{conf.}} <|f(\beta)|
\label{eq:oldconf}
\end{equation}
This works very well in the Gaussian case because $f(\beta)$ is never zero. On the 
other hand for a non-Gaussian distributions with complex zeros, this criteria will 
exclude a neighborhood of the zero. 

\begin{figure}
\includegraphics[width=0.8\columnwidth]{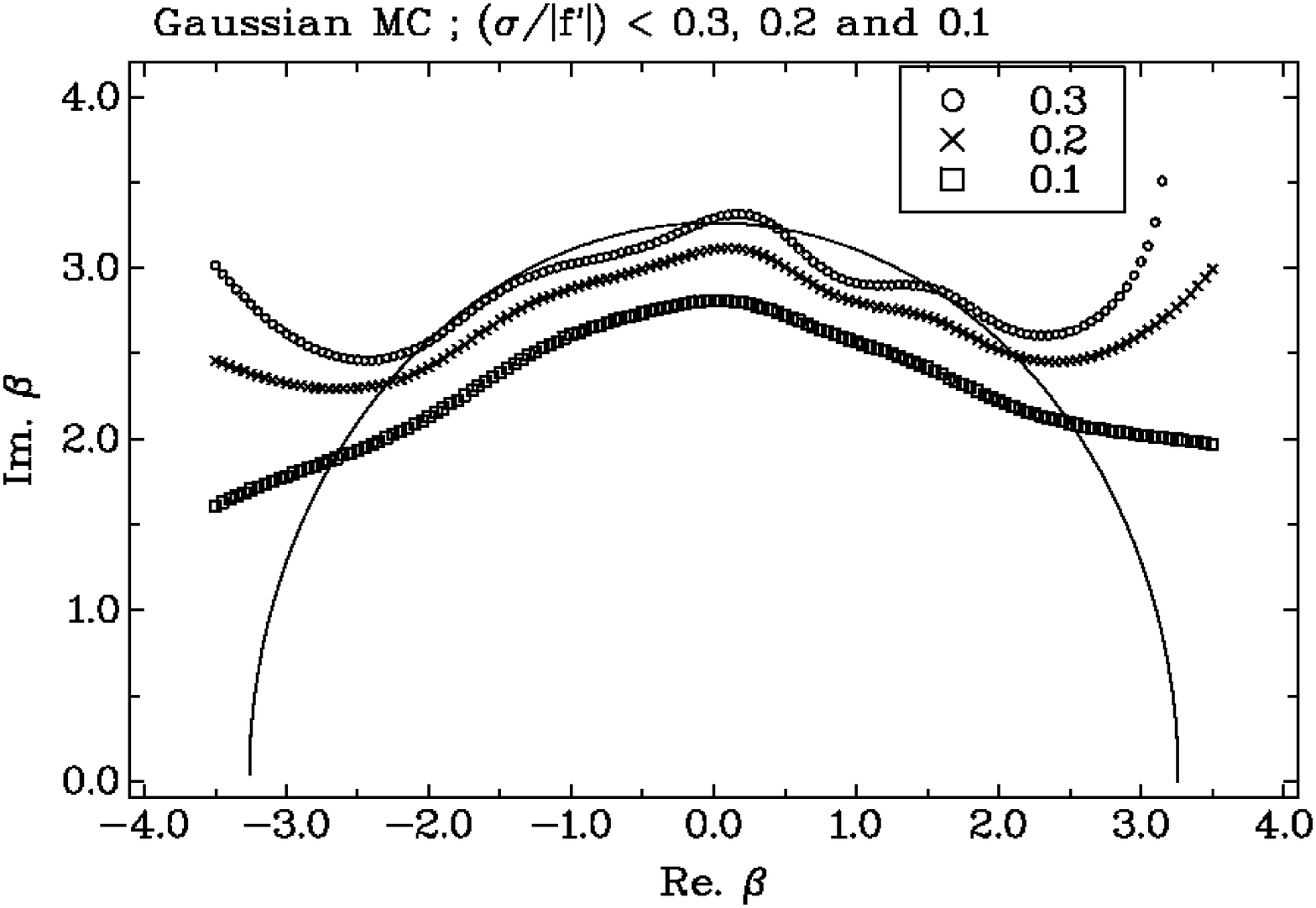}
\caption{Boundary of the confidence region for $d$ = 0.3 (circles, upper set), 0.2 (crosses, middle set) and 0.1 (boxes, lower set), compared to the Gaussian circle of confidence, all for 40,000 configurations. }
\label{fig:newG}
\end{figure}

We would like to know how the MC level curves move as a consequence of the uncertainty in 
$f$. A possible method to search for the zeros of say the real part consists in following this quantity in a particular direction and look for the change of sign. 
The error on the location of the zero is then the error on the function divided by the slope. This error depends on the direction. It can be minimized by moving in the normal 
direction to the level curve where the slope is maximal. A complex generalization of this estimate is the ratio $\sigma _f(\beta) /|f'(\beta)|$. This defines direction 
independent circles of uncertainty around every point on the complex $\beta$ plane. 
We propose to consider the alternative regions of confidence defined by the 
condition.
\begin{equation}
\frac{\sigma _f }{\sqrt{N_{conf.}}\ |f'(\beta)|}<d\ .
\label{eq:newconf}
\end{equation}
In order to be useful $d$ should be a fraction of the typical distance between zero level curves of the real and imaginary part. Note that for the Gaussian distribution, $f_G(\beta)=\exp[(1/2)\beta^2]$ and consequently $f'_G=\beta f_G$. If we impose that the 
the two conditions should be equivalent on the original circle of confidence Eq. (\ref{eq:oldconf}), we obtain $d\simeq 0.3$ for $N=40,000$. We have checked 
(Fig. \ref{fig:newG}) that this 
condition produces a region of confidence very close to the Gaussian circle of confidence for $|\beta|<2$. We have seen in Sec. \ref{sec:error}, that for $|\beta|>2$ 
we usually underestimate the variance and the region of confidence becomes larger than it should be. 

Note that Eq. (\ref{eq:oldconf}) guarantees at 84 percent confidence 
level that we do not have artificial zeros. For the purpose of establishing lower bounds, one might want have a higher confidence level. The same remarks apply for the 
new criterion. Empirically, we found that $d=0.2$ removes an upper sliver of the Gaussian 
confidence circle where most of the occasional artificial zeros appear. 
This is illustrated in Fig. \ref{fig:newG}.

A practical remark. In order to calculate $|f'(\beta)|$, we can make use of the 
fact that $f$ is an analytical function. This is obviously true in the Gaussian case. 
This is also obviously true for any average over a finite number of configurations. 
Consequently, we may use the Cauchy-Riemann conditions. 

Assuming analyticity, one can calculate the derivative in an arbitrary complex direction.
For instance, if we search for zeros at fixed values of $Re \beta$, it is convenient 
to use the derivative in the imaginary direction. In this particular case, we have 
\begin{equation}
	|f'|=\sqrt{(\partial R/\partial y)^2+(\partial I/\partial y)^2}
\end{equation}

\section{Examples of quasi-Gaussian distributions}
\label{sec:exs}
\def\pt{\tilde{P}}
\def\st{\tilde{S}}
\def\lt{\tilde{\lambda}}
\def\bt{\tilde{\beta}}
We now consider normal Gaussian distribution with small perturbations of order 3 and 4:
\begin{equation}
\pt(\st)\propto \exp(-(1/2)\st^2-\lt_3 \st^3-\lt_4 \st^4)
\end{equation}
Note that due to the non-Gaussian corrections, $<\st>\neq 0$ and the variance is usually not 1. For this reason, we use the tilted symbol to emphasize that the transformation Eq. (\ref{eq:can}) needs to be used in order to have the natural units 
advocated in Sec. \ref{sec:natural}. 
After this transformation is performed we get 
\begin{eqnarray}
P(S)&\propto&\pt(<\st>+\sigma_{\st}S)\nonumber \\
&\propto& \exp(-\lambda_1 S-\lambda_2S^2-\lambda_3 S^3-\lambda_4 S^4)
\label{eq:ps}
\end{eqnarray}
where the untilded symbol corresponds to the reduced units (but the ``red.''subscript  is dropped). 
\subsection{Example 1: $\lt_3=0.1,\ \lt_4=0.01$ }
\begin{figure}
\includegraphics[width=0.8\columnwidth]{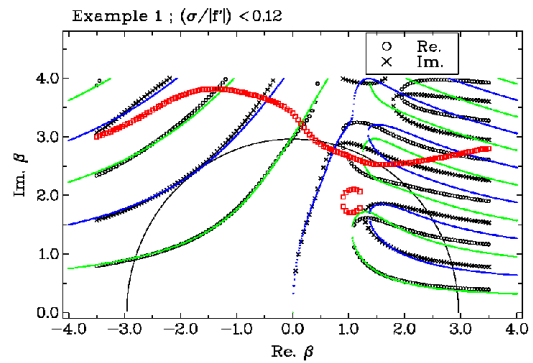}
\caption{Zeros of the real (circles) and imaginary (crosses) part for 
40,000 configurations corresponding to the first example. The small dots are the 
accurate values for the real (gray, green on-line) and imaginary (black, blue online) parts. The exclusion region boundary for $d=0.12$ is represented by boxes (red online). 
The solid line is the circle of confidence of the Gaussian approximation.}
\label{fig:ex1z}
\end{figure}

We first consider the case $\lt_3=0.1,\ \lt_4=0.01$. It has been chosen in such a way 
that we have zeros inside and outside the Gaussian region of confidence. 
It is possible to find the zero level curves of the real and imaginary parts by 
using numerical integration. We refer to these curves and to the complex zeros where they intersect as  ``accurate''.
The accurate zero level curves are compared with the MC ones in Fig. \ref{fig:ex1z}. 
The accurate complex zeros are located at $\beta=1.3735+i1.7104$ and $\beta=1.3735+i2.9478$. As far as we can tell the two real parts are the same. For further 
reference, the variance of the original variable $\st$ is 1.0988 and the location of the 
real part of the complex zeros is 1.2500 in unrescaled units. The MC zeros closest to these accurate zeros are located near 1.3+i1.6 (close to the first one) and 0.9+i2.8 (not so close to the second one). We have considered several values of $d$ to define a region of confidence. 
It clear that it should contain the lowest zero which is a good approximation. On the 
other hand, the second zero is produced by curves that depart significantly from the 
accurate ones. A value of $d=0.12$ defines a confidence region that contains the lowest 
MC zero but unambiguously excludes the other zero (see Fig. \ref{fig:ex1z}). 
Note that there is a small island of exclusion not so far from the lowest zero.
This is in agreement with larger departures where the curves turn over. When $d$ is 
increased, the upper boundary moves up and the island shrinks. For $d>0.2$, the confidence region includes the second zero. 

\subsection{Example 2: $\lt_3=0.01,\ \lt_4=0.002$}
\begin{figure}
\includegraphics[width=0.8\columnwidth]{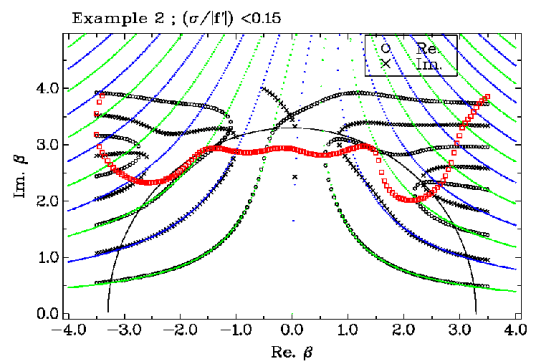}
\caption{Zeros of the real (circles) and imaginary (crosses) part for 
40,000 configurations corresponding to the second example. The small dots are the 
accurate values for the real (gray, green on-line) and imaginary (black, blue online) parts. The exclusion region boundary for $d=0.15$ is represented by boxes (red online). 
The solid line is the circle of confidence of the Gaussian approximation. }
\label{fig:ex2z}
\end{figure}

Our second example is $\lt_3=0.01,\ \lt_4=0.002$. The perturbation is much smaller and 
the first accurate zero is at $\beta=1.207+i5.241$, far outside the Gaussian circle of 
confidence. All the zeros seen in Fig. \ref{fig:ex2z} are artifacts. The value $d=0.15$ 
excludes all them with a good margin of safety. 

\section{Conjecture concerning the real part of the complex zeros}
\label{sec:conjecture}
In Sec. \ref{sec:error}, we mentioned the possibility of probing the tails of the 
distribution by introducing a bias. In this section, we will explore this possibility 
by introducing a real parameter $b$ to favor one side of the distribution:
\begin{equation}
	\pt_b(\st)=\pt(\st)\exp(-b\st)\ .
\end{equation}
In the language of the original problem it amounts to shift the value of $\beta_0$. 
In this section, we do {\it not} perform the change of variables as in Eq. (\ref{eq:can}), because we want to discuss the $b$ dependence of the variance 
(which would become 1 after the rescaling). For this reason we keep the tilde in the 
equations. If we denote the average with the new distribution as $<...>_b$, we have 
the simple identity
\begin{equation}
	<\exp(-\bt \st)>_0=C(b)<\exp(-(\bt-b)\st)>_b\ .
\end{equation}
$C(b)$ is a normalization factor that will be assumed non-zero. This assumption 
can be justified as long as there are no zeros of the partition function on the real axis (we remind that $b$ is real). It is then clear that the effect of the bias is to 
shift a complex zero by -$b$. 

\begin{figure}
\includegraphics[width=0.8\columnwidth]{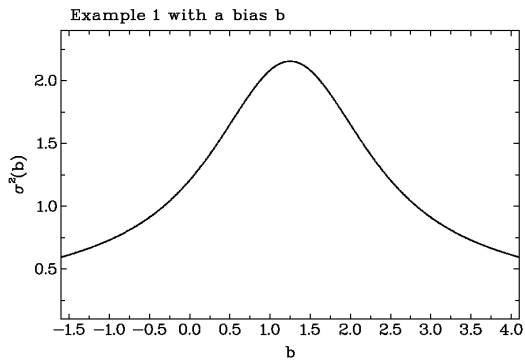}
\caption{Second moment as a function of $b$. }
\label{fig:var}
\end{figure}

In the Gaussian case, the zeros have a mirror symmetry with respect to the imaginary axis.
When odd perturbations are introduced, the symmetry is broken. In Figs. \ref{fig:ex1z} and \ref{fig:ex2z}, the complex zeros are on the right and the curves are in some sense 
tilted to the right. As we increase $b$ by positive values, the complex zeros move to the left.
The third moment of the distribution, 
$<(\st-<\st>)^3>_b$ can be seen as a measure of the left-right asymmetry. 
We conjecture that when the real part becomes zero, the third moment vanishes, which corresponds to an extremum of the second moment. This conjecture has been checked 
very precisely with the two examples of Sec. \ref{sec:exs}. It is illustrated in Fig. 
\ref{fig:var} for example 1. We see that the second moment peak near 1.25 which 
coincides with the location of the real part (before rescaling) of the complex zeros.  

If this conjecture remains true even approximately, it is telling us to look for 
complex zeros for real values on a vertical strip that includes the region(s) of the 
real axis where the second moment is maximum.

\section{Indirect methods to locate the zeros}
\label{sec:indirect}

So far, we have not discussed methods to find complex zeros with a large imaginary 
part. This may occur as in Example 2 when the corrections to the Gaussian form are 
small. Under these circumstances, the third and fourth moments, that would vanish in 
the Gaussian case, are possibly very small and it might be possible to infer the 
small non-Gaussian corrections from these small moments. For completeness, the third and fourth moments are defined as 

\begin{eqnarray}
m_3&=&<S_{red.}^3>\nonumber\\
m_4&=&<S_{red.}^4>-3<S_{red.}^2>^2
\end{eqnarray}

We have used Newton's method to determine the unknown parameters $\lambda_1, \dots \lambda_4$ in Eq. (\ref{eq:ps}) from the first four moments calculated with MC. 
This is the only input and it can be calculated easily for more complicated models. 
The numerical estimates of the moments are based on a single set of 40,000 configurations obtained as described above. 
We worked with the 
reduced variables so the first two moments are fixed to 0 and 1. The integrals and 
the derivatives in Newton's method were calculated numerically. In example 2, the corrections are small 
and the initial values (0,1/2,0,0) corresponding to the normal Gaussian is a good approximation. Newton's method converges rapidly for these initial values. 
For example 1, the corrections are larger and one needs to start closer to the solution. The main source of uncertainty come from errors in the determination of the 
moments from MC data. 
The numerical results are shown in Tables 1 to 4. 

\begin{table}[t]
\begin{tabular}{||c|c|c||}
\hline
&MC&Accurate\cr
 \hline
$m_3$ & -0.575&-0.559\cr 
$m_4$& 0.498&0.448  \cr 
  
\hline
\end{tabular}
\caption{Third and fourth moments from MC and accurate numerical integration in example 1.}
\end{table}
 
 \begin{table}[t]
\begin{tabular}{||c|c|c||}
\hline
& MC+Newton &Accurate\cr
 \hline
$\lambda_1$&-0.320406 & -0.315192\cr 
$\lambda_2$& 0.50140&0.496627 \cr 
$\lambda_3$&0.117795 & 0.115924\cr 
$\lambda_4$&0.01433 &0.014577 \cr  
 \hline
\end{tabular}
\caption{$\lambda_i$ from Newton's method compared to their accurate values in example 1.}
\end{table}

 \begin{table}[t]
\begin{tabular}{||c|c|c|c||}
\hline
&MC I&MC II&Accurate\cr
 \hline
$m_3$ & -0.0469&-0.0576&-0.0535\cr 
$m_4$&-0.0408&-0.0371&-0.0353\cr 
  
\hline
\end{tabular}
\caption{Third and fourth moments from MC I (40,000 data points), MC II (1,600,000 data points) and accurate numerical integration in example 2.}
\end{table}
 
 \begin{table}[t]
 \label{table:lambda}
\begin{tabular}{||c|c|c|c|c||}
\hline
& MC I+Newt.&MC II+Newt.&Min. $\chi^2$  &Accurate\cr
 \hline
$\lambda_1$& -0.0247&-0.0303  &-0.0299&-0.0280493\cr 
$\lambda_2$&0.488& 0.489 &0.490&0.489364\cr 
$\lambda_3$&0.00837&0.0103 &0.0101& 0.00948691\cr 
$\lambda_4$&0.00209&0.00206 &0.00189&0.00192227\cr  
  
  \hline
\end{tabular}
\caption{$\lambda_i$ from Newton's method with the two sets of MC data described in the text, and from th minimization of $\chi^2$ compared to their accurate values in example 2.}
\end{table}

\begin{figure}
\includegraphics[width=0.8\columnwidth]{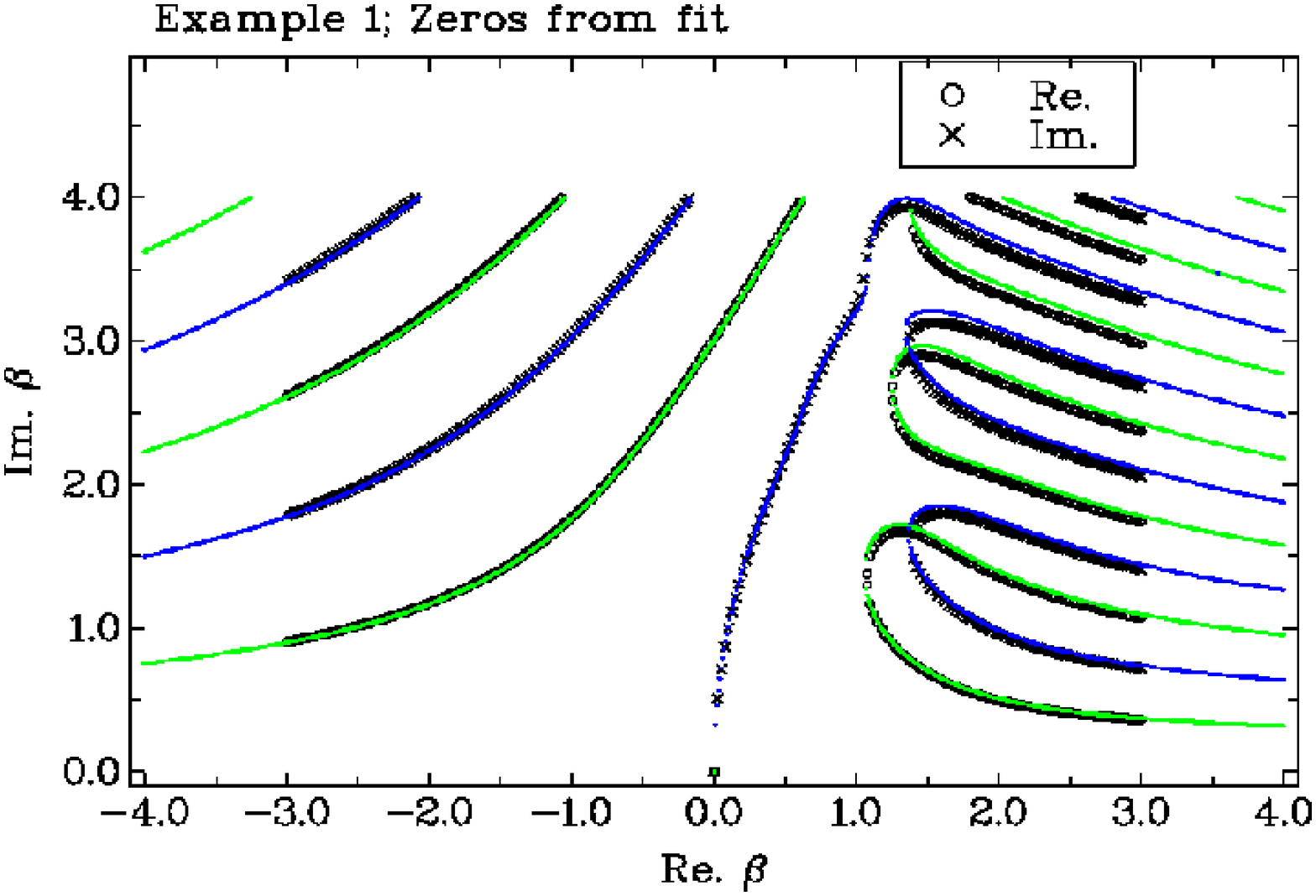}
\includegraphics[width=0.8\columnwidth]{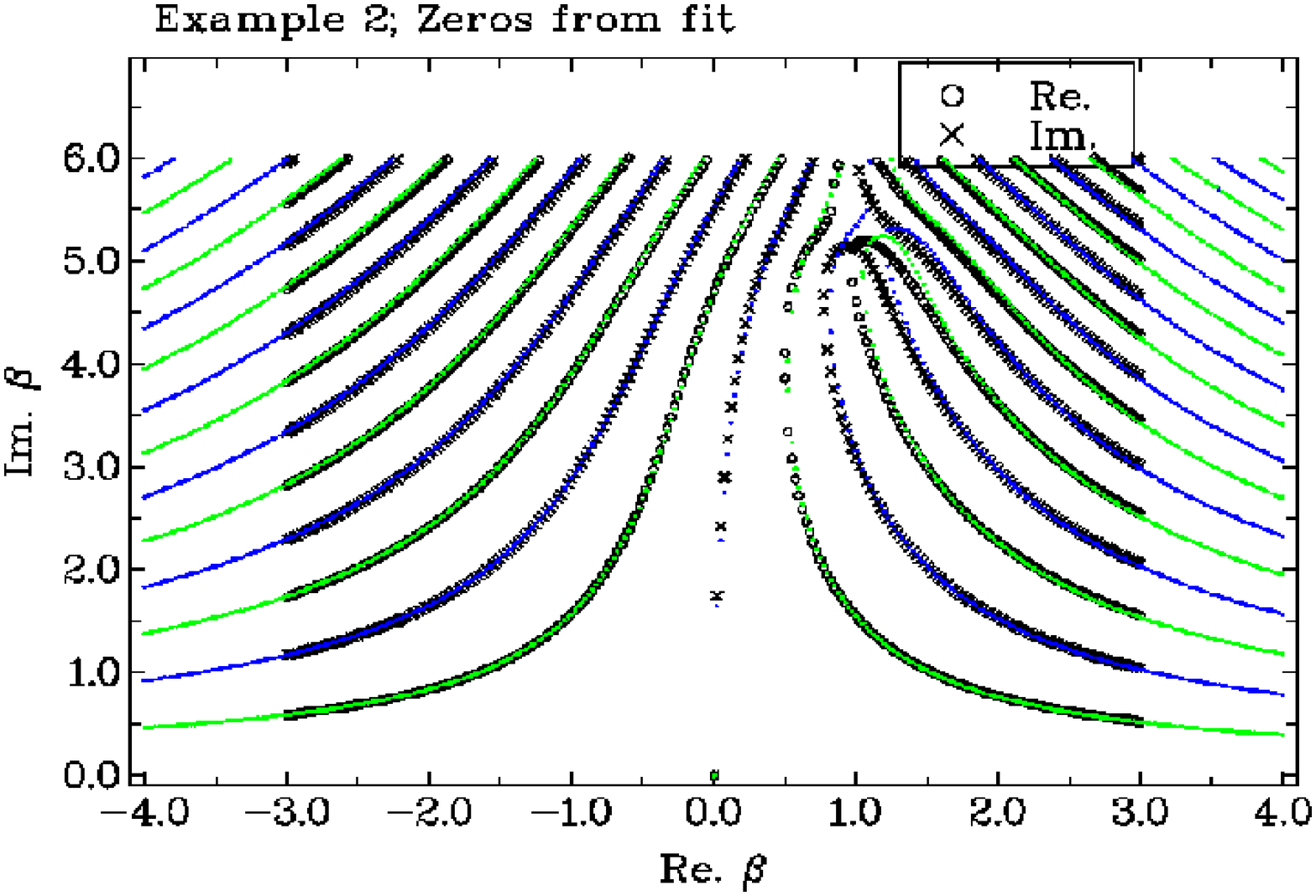}
\caption{Zeros of the real (circles) and imaginary (crosses) using the approximate $\lambda_i$ obtained from MC moments. The small dots are the 
accurate values for the real (gray, green on-line) and imaginary (black, blue online) parts, for example 1 (upper graph) and example 2 (lower graph). }
\label{fig:zfit}
\end{figure}

Using the approximate values of $\lambda_i$ obtained from the MC moments and Newton's 
method, we were able to construct approximate zero level curve for the real and imaginary parts the results are shown in Fig. \ref{fig:zfit}. In example 1, the agreement between the two sets of curves is very good and allows an accurate 
determination of the first two zeros. In example 2, the agreement is also very good  except near the complex zero where the discrepancy is visible. The value of the 
complex zero obtained with the new method is 0.98+i5.14 which differs from 
the accurate value 1.207+i5.241. We have repeated the same calculation with other 
sets of 40,000 configurations obtained with the bootstrap method and obtained similar size for the discrepancies. 

We have tried to improve the accuracy of the estimates. First, we have used the entire 
data (1,600,000 values) to determine the moments. This improved the accuracy of the 
moments (see Table III) and of the parameter $\lambda_i$ (Table IV). The complex 
zero was located at 1.21+i5.08. We have also used a minimization of the $\chi^2$ of the 
histogram to determine the $\lambda_i$. The results are very good (see Table IV) and 
produces level curves that a almost impossible to distinguish from the accurate ones on a graph.

As the perturbations get smaller, the zeros get a larger imaginary part and the numerical integration becomes more difficult because of the fast oscillations of the
integrand. However, it possible to use perturbative methods. When $\lambda_3$ and $\lambda_4$ are both zero, the problem is Gaussian, solvable analytically and there are no complex zeros. If we calculate $<\exp(-\beta S)>$ at first order 
in $\lambda_3$ and $\lambda_4$ and divide by the Gaussian limit (which has no zeros), 
we obtain a polynomial of order 4 in $\beta$:
\begin{eqnarray}
	&<&\exp(-\beta S)(1-\lambda_3 S^3-\lambda_4 S^4)>_G/<\exp(-\beta S)>_G\cr \nonumber&=& Q(\beta)\cr &=& 1+\dots -
	\lambda_4 \beta^4/(16\lambda_2^4)\ ,
\end{eqnarray}
where $<\dots>_G$ is a notation for the average with  a Gaussian weight as in Eq. (\ref{eq:ps}) with $\lambda_3$ and $\lambda_4$ set to zero. 
After expansion, the polynomial $Q(\beta)$ has 16 terms which can be calculated 
easily by Gaussian integration. For the accurate values of $\lambda_i$  of Table 
\ref{table:lambda}, we have zeros of $Q(\beta)=0$ for $\beta=1.08\pm i4.56$ which is not 
far from the correct value.
\section{Conclusions}
We have build a ''ladder'' of methods that can be applied for increasing values of the 
imaginary part in natural units. We have refined the determination of the region where MC calculation can be trusted. 
Fitting methods based on cubic and quartic perturbations work for larger values of the 
imaginary part. It is clear that the parametric form of the distribution is known 
in the two examples considered here. When the form is not known, this introduces an additional uncertainty. 
The zeros for the approximate distributions can be found 
using numerical integration provided that the imaginary part is not too large (about 5 to 6 in natural units). Larger values indicate small non-Gaussian terms and can be treated by perturbative methods. These methods are being experimented on $SU(2)$ and $SU(3)$ LGT. This will be reported in a subsequent publication.

\begin{acknowledgments}
This 
research was supported in part  by the Department of Energy
under Contract No. FG02-91ER40664 and the National Science Foundation 
NSF-PHY-0555693.
\end{acknowledgments}

\end{document}